\documentstyle[12pt]{article}

\def\beq{\begin{equation}}
\def\eeq{\end{equation}}
\def\beqa{\begin{eqnarray}}
\def\eeqa{\end{eqnarray}}

\newcommand{\sect}[1]{\setcounter{equation}{0}\section{#1}}

\newcommand{\EQ}{\begin{equation}}
\newcommand{\EN}{\end{equation}}
\newcommand{\bea}{\begin{eqnarray}}
\newcommand{\ena}{\end{eqnarray}}

\begin{document}

\vskip 1.2cm

\centerline{\Large \bf A unique theory of all forces}

\vskip 1.2cm

\centerline{\bf Paolo Di Vecchia}
\centerline{\sl NORDITA}
\centerline{\sl Blegdamsvej 17, DK-2100 Copenhagen \O, Denmark}
\vspace{1cm}
\centerline{ {\it Talk given at the Conference ''Beyond the Standard Model" , V
}}
\centerline{\it{Balholm, Norway  (May 1997)}}
\vspace{1cm}
%
%
%
%
%
%

\begin{abstract}
In discussing the construction of a consistent theory of quantum gravity
unified with the gauge interactions we are naturally led to a string theory.
We review its properties and the five consistent supersymmetric string theories
in ten dimensions. We finally discuss the evidence that these theories are
actually special limits of a unique 11-dimensional theory, called M-theory,
and a recent conjecture for its explicit formulation as a supersymmetric 
Matrix theory.
\end{abstract}

\vskip 1cm
%
\sect{Unification of all interactions}
\label{intro}
\vskip 0.5cm

The electro-weak and strong interactions are described by the
so-called standard model of particle physics. It is a gauge field theory based
on the direct product of the group $SU(3)$ colour with coupling constant 
$g_3$ for the strong interactions and the group $SU(2) \otimes U(1)$ with 
coupling constants $g_2$ and $g_1$ respectively for the electro-weak 
interactions.   
A non abelian gauge theory is a generalization of quantum 
electrodynamics (QED) to 
the case of a non-abelian gauge group  and is described by a Lagrangian of 
the type:
\beq
{\cal{L}} = - \frac{1}{4 g^2} F_{\mu \nu}^2 + {\bar{\Psi}} \left( i 
\gamma^{\mu}{D}_{\mu} - m \right) \Psi \hspace{1cm}; \hspace{1cm}  
D_{\mu} = \partial_{\mu} + i A_{\mu}
\label{lag}
\eeq
The field $\Psi$ describes the matter fields that in the standard model are
the quarks and leptons, while the gauge field $A_{\mu}$ and its field 
strenght $F_{\mu \nu}$
describe the gauge particles responsible for providing the interaction among
the matter particles. In the standard model the gauge fields are the $8$ 
gluons for the strong interactions and the $W^{\pm}$, the $Z^0$ and the 
photon for the electro-weak interactions.     

When one quantizes $QED$ or the standard model one finds infinities that are
the reflection of the infinities that are already present at the classical 
level. Remember that, for instance, already in classical 
electromagnetism~\cite{LANDAU} one 
must introduce a classical radius $r_0$ for the electron, in order to avoid 
short distances divergences directly related to the 
\underline{pointlike nature} of the electron, defined by the 
$e{^2}/{r_0} = m c^2$.
These infinities persist in the quantum theory and must be renormalized 
away in order to obtain finite quantities to be compared with the experiments. 
One of the most important consequences of the renormalization theory is the
fact that the coupling constants are not fixed once and for all as in the 
classical theory, but are running with the energy at which we perform the
experiments. Their running   is completely fixed by the 
particle spectrum at low energy and by their quantum numbers under the gauge 
group of the standard model, at least up to an energy at which new physics and 
new particles appear. Since the three gauge coupling constants of the 
standard model have been measured with very high precision at LEP at an energy
equal to the mass of the $Z^0$ gauge boson and one knows the particle spectrum
of the standard model one can see if they will meet at high energy. 
It turns out that they do not  meet at a common value in the case of the 
standard model~\cite{AMALDI}, but, if we extend 
the standard model by introducing for each
particle of the standard model a supersymmetric partner obtaining the spectrum
of the minimal supersymmetric standard model, then one can choose the 
supersymmetric breaking scale comparable with  the electro-weak scale ($10^2,
10^3 \, GeV$) in such a way that the three couplings now meet in a single 
point~\cite{AMALDI}, that we call $\alpha_{GUT}$, at an energy 
$M_{GUT}$. 
The values found for these two quantities at the unification point  are
\beq
M_{GUT} = 2 \cdot 10^{16} GeV \hspace{2cm} \alpha_{GUT} \equiv 
\frac{g_{GUT}^{2}}{4 \pi}  \sim \frac{1}{25}
\label{gut}
\eeq
Supersymmetry is not only consistent with the fact that the three couplings
unify at high energy, but it is also required for other reasons as for instance
the hierarchy problem. Because of them it is believed, but far from 
proved, that at an energy of the order of the electro-weak scale the standard 
model has to be modified to include supersymmetry. Supersymmetry generalizes 
the Lorentz invariance of  special relativity theory and the translational 
invariance responsible for the energy conservation, in a symmetry where bosons 
and fermions coexist in the same multiplets. 

The previous considerations imply that at sufficiently high energy also 
the strong interactions will be unified with the electro-weak interactions.
This is, of course, a very appealing idea, but at this point we must not 
forget that in nature
we observe another kind of interaction, the gravitational interaction. 
The gravitational interactions are described by the Einstein's theory of
general relativity in which the gravitational force is produced by the fact
that the space-time is curved. The action for the 
gravitational field is given by the Einstein's action:
\beq
S = \frac{c^3}{16 \pi G_{N}} \int d^4 x \sqrt{-g} R
\label{genrel}
\eeq
where $R$ is the scalar curvature of the space-time and $G_N$ is the Newton
constant that in the newtonian limit enters in the gravitational force $F$
between two equal masses $M$ at a distance $R$ given by $F = - 
G_{N} M^{2}/{R^2}$.
The gravitational attraction between two masses becomes strong when the 
dimensionless coupling constant $G_N M^2 / \hbar c \sim 1$. This equality
defines the Planck mass given by
\beq
M_P = \sqrt{ \frac{\hbar c}{G_N}} = 1.2 \cdot 10^{19} GeV
\label{planckmass}
\eeq
 
When we approach energies of the order of the Planck mass then also gravity
must be quantized, but Einstein's theory of general relativity is not
a renormalizable theory as are gauge theories. This is due to the fact that
a gauge field couples to the charge, while gravity couples to the energy
giving extra positive powers of the momenta circulating in the loops and 
making the loop integration more divergent so that the theory becomes
non-renormalizable. In other words in the case of gravity the pointlike 
structure of
field theory creates such divergences that the theory cannot be given a
quantum meaning at least in perturbation theory.

We face therefore the problem of defining a quantum theory of gravity. 
Another question has to do with the unification that we have seen between
the strong and the electro-weak interactions at a scale that is not so
different from the Planck scale. If the strong and the electro-weak 
interactions are really unified at a unification scale that is only a factor 
$10^3$ smaller than the Planck scale why not to try to unify all interactions 
at a sufficiently high energy?  How this can be done is discussed 
in the next section.

\sect{From point particles to strings}
\label{strings}
\vskip 0.5cm
We have previously stressed that the ultraviolet infinities of the various
field theories are due to the fact that they describe pointlike objects. 
In the case of gravity the short distance divergences are so strong that,
if we want to construct a quantum theory of gravity, we are obliged to
abandon the idea that the fundamental constituents of nature are pointlike
objects. The simplest possible extension is of course that they are 
tiny one-dimensional strings. 

The string model originated at the end of the
sixties as an attempt to describe, in the framework of  $S$-matrix theory, 
the physics of strong interactions with an infinite set of resonances, 
but it became soon clear that it incorporated also massless gauge fields and 
gravitons that are not present in the low energy hadron spectrum.  
It predicted also an exponential decrease of the cross sections at large 
transverse momentum in 
contradiction with the experiments which showed instead only a power decrease 
typical of pointlike structures. These unphysical features of  string 
model brought back the research on strong interactions in the realm of field 
theory and only a few years later the theory of quantum chromodynamics (QCD) 
was formulated as the one correctly describing the strong interactions. 

On the other hand, exactly for those features that were in disagreement with 
the physics of strong interactions, it was soon recognized that the string 
model could provide 
a theory in which all interactions  including gravity were unified~\cite{SS}. 
In particular the exponential cutoff at high momentum was an essential feature 
of the model for providing a finite quantum theory of gravity. 

As a spinless point particle is described by its position $x^{\mu} (\tau)$
in  Minkowski space-time as a function of its proper time $\tau$, so a
bosonic
string is described by its position $x^{\mu} (\tau, \sigma)$ as a function
of $\tau$ and of an additional variable $\sigma$ that parametrizes the 
various points of
a  string. In going from zero-dimensional to one-dimensional objects the
free action for a particle becomes that of a string:
\beq
S_{p} = - \frac{mc}{2} \int (\frac{\partial x}{\partial \tau})^2 
d \tau \Longrightarrow
S_{s} = - \frac{T c}{2} \int d \tau \int d\sigma \left[ 
\left( \frac{\partial x}{\partial \tau} \right)^2 - 
\left( \frac{\partial x}{\partial \sigma} \right)^2 \right]  
\label{action}
\eeq
where $m$ is the mass of the particle, while $T$ has the dimension of an 
energy per unit lenght and it is called the string tension.

If a particle has also non zero spin for instance equal to $1/2$ we need to
introduce together with its position in the space-time $x^{\mu} (\tau)$ also
a Grassmann coordinate $\psi^{\mu} (\tau )$ describing its spin degrees of
freedom. In the same way a string having also spin $1/2$ degrees of freedom
distributed along the string is also described by its Minkowski coordinate
$x^{\mu} (\tau, \sigma)$ together with a Grassmann coordinate $\psi^{\mu}
( \tau, \sigma )$. Such a string is called a superstring. One can also have
internal degrees of freedom distributed along the string. In this case one
has a heterotic string. In addition strings can be both open and closed. 

Classically one can find the most general motion of a string and in general
the dynamics of a string can be understood as arising from the balance between
the centrifugal force that will tend to make the string longer and the force
due to the string tension that will tend to push a string to become a point.

When a string is quantized one finds an infinite spectrum of states lying on
linearly rising Regge trajectories with slope $\alpha'$ related to the string
tension through the relation $T = [ 2 \pi \alpha ']^{-1}$. In particular the
spectrum contains a massless sector with spin $1$ gauge fields and a spin $2$
graviton showing that the theory is able to unify gauge interactions with 
Einstein's theory of general relativity. In addition in the case of 
superstring and heterotic string, that are the only theories free from 
infrared problems as for instance tachyons and therefore the only fully 
consistent ones, one finds that they are supersymmetric. It is actually 
this symmetry that saves them from the inconsistencies of the bosonic string.
Therefore the existence of a new symmetry between
bosons and fermions, called supersymmetry, is a prediction of string theory.

A string theory has a huge spectrum of massive states with masses and
degeneracy given respectively by $M^2 = N/\alpha '$ and by $ d(M) \equiv 
e^{ c \sqrt{\alpha '} M}$,
where $c$ is a number depending on the particular string theory and $N$ is
a non negative integer. 

In the field theory limit corresponding to the infinite tension limit ( $ T
\rightarrow \infty$, $\alpha ' \rightarrow 0$), that is the limit where a 
string becomes a particle because, in this limit, the centrifugal force is 
not able anymore to compensate the string
tension, we see that only the massless  states survive in the spectrum, while
all the massive states acquire an infinite mass and disappear from it.

Strings can also interact, but, unlike  pointlike particles
where the interaction is obtained by introducing interaction vertices and the
interaction point is a point of singularity, no interaction vertex is needed
for strings and the world tubes of strings join smoothly. 

It turns out that superstrings and heterotic strings cannot be consistently
quantized unless the space-time dimension  of  Minkowski space $D \leq 10$.
In particular the supersymmetric strings  live naturally at $D=10$
where one finds five fully consistent and inequivalent string theories.
They are all supersymmetric, four of them are theories of closed strings,
while one is a theory of both open and closed strings. Two of them do not
include gauge interactions, while the other three include them but with only
two very special gauge groups: either $SO(32)$ or $E_8 \otimes E_8$.
In order to make contact with the observed phenomenology at presently 
available energies one must assume that $6$ of the $10$ dimensions are not
expanding as the other $4$. They remained small, highly curved and 
compactified {\`{a}} la Kaluza-Klein. 

Concerning the unification
of the gauge couplings, string theory is more powerful than field theory
because now also gravity is fully unified with the gauge interactions. In 
particular, in the heterotic string at  tree level at the unification scale
one finds~\cite{GINSPARG} a relation
between the Newton's constant $G_N$, the string coupling constant $g_s$ and
the gauge coupling constants of the various gauge groups $g_i$'s given by:
\beq
\frac{8 \pi G_N}{\alpha '} = g_{i}^{2} k_i = g_{s}^{2}
\label{struni}
\eeq
where the constants $k_i$ are the central charges of the Kac-Moody algebras.
When one takes into account  the one loop corrections one can see that the
gauge coupling constants will run according to the renormalization group and
the scale of the grand unification can be computed in terms of the string 
coupling constant and the Planck mass obtaining~\cite{KAPLU} ($\gamma$ is the
Euler-Mascheroni constant):
\beq
M_{GUT} = \frac{e^{(1- \gamma)/2} 3^{-3/4}}{4 \pi} g_s M_{P} = 5.27 \cdot 
10^{17} g_s \,\, GeV
\label{unimass}
\eeq
But, since $g_s \sim 1$, as follows from the second equation in eq. 
(\ref{gut})
where $g_s$ must be identified with $g_{GUT}$, one gets a discrepancy of a
factor $20 \sim 30$ between eq. (\ref{unimass}) and the first equation in 
eq. (\ref{gut}). This discrepancy is independent on the specific 
compactification of the six extra dimensions and goes under the name of the
string gauge coupling unification problem. In conclusion, although one 
apriori could have expected that the 
unification scale of the gauge and gravitational interactions should have been
of the order of the Planck mass, it turns out that such a scale is clearly 
smaller than $M_P$ as follows from eq. (\ref{unimass}), but, however, still a 
factor $20 \sim 30$ too high in comparison with the extrapolation from the 
low energy experiments. The various possible resolutions of this problem are
reviewed in Ref.~\cite{DIENES}.     

We have seen that, in order to make contact with the observed phenomenology,
one must compactify six of the ten dimensions.
They can be compactified in various ways obtaining a huge number of 
four-dimensional consistent string 
theories. Such huge arbitrariness seemed in contradiction with the uniqueness
character that a unified theory of all forces is supposed to possess.
Although we now understand that the huge amount of arbitrariness is just a 
consequence of the many possible ways in which we can Kaluza-Klein 
compactify the ten-dimensional consistent theories, it 
is, however, 
still unpleasant to have to accept the existence in ten dimensions of five 
consistent and inequivalent theories.  
Why five and not just one if this is the fundamental theory of all 
interactions?

\sect{Non-perturbative equivalence and $M$-theory}
\label{nonper}
\vskip 0.5cm

In order to be able to answer the previous question that has puzzled string 
theorists for more than ten years, we must 
step back for a second, go back to field theory and see if it is
conceivable there that two theories, that are very different when we study them
order by order in their respective perturbation theories, 
can be actually completely equivalent if we analyze them with non-perturbative
methods. Actually it turns out that this possibility is not only abstractly 
open in field theory, but is also realized in some systems in two dimensions 
as in the Ising model or in the case of the quantum equivalence between the 
sine-Gordon and Thirring theory~\cite{COLE}. 

If we consider the sine-Gordon theory described by the two-dimensional 
Lagrangian:
\beq
L_{Sine-Gordon} = \frac{1}{2} \left( \partial_{\mu} \Phi \right)^2 - 
\frac{M^2}{\beta} \cos \beta \Phi
\label{sineg}
\eeq
and the Thirring theory described by
\beq
L_{Thirring} = {\bar{\Psi}} \left( \gamma^{\mu} \partial_{\mu} - m \right) 
\Psi + \frac{g}{2} \left(  {\bar{\Psi}} \gamma^{\mu} \Psi \right)^2
\label{Thirring}
\eeq
it has been shown~\cite{COLE} that, although they look very different and one 
is a bosonic theory while the other is a fermionic one, they are actually  
equivalent provided that the two coupling constants are related through the
relation:
\beq
\beta^2 = \frac{4 \pi}{1 + g/ \pi}
\label{equi}
\eeq
As a consequence strong coupling in the Thirring model implies weak
coupling in the sine-Gordon theory and viceversa. It can also be seen that 
the sine-Gordon theory has soliton solutions of the classical equations of
motion corresponding in the quantum theory to particles with a mass 
proportional to the square of the inverse of the sine-Gordon coupling
constant $\beta$. They are not appearing as fields in the sine-Gordon 
Lagrangian, but are in fact described in the Thirring theory by the 
fundamental field $\Psi$.

Another example is the 
two-dimensional Ising model where the partition function $Z(K)$ computed
on a square two-dimensional lattice 
\beq
Z(K) \equiv \sum_{\sigma= \pm 1} e^{ K \sum_{<ij>} \sigma_i \sigma_j} 
\hspace{2cm};\hspace{1cm} K = \frac{J}{K_B T}
\label{ISI}
\eeq
turns out to be  equal to the one computed 
on the dual lattice obtained from the original square lattice by taking the
points situated at the center of the squares of the lattice:
\beq
Z(K) = Z^{*}( K^{*})  \hspace{2cm} {\rm{if}} \hspace{2cm} \sinh(2 K^{*}) = 
\frac{1}{\sinh(2 K)}
\label{equiva}
\eeq
where $K_B$ is the Boltzmann constant, $T$ is the temperature and $J$ is the
coupling constant between two next neighbouring lattice points. The
partition function $Z(K)$ gives a good perturbative description of the 
high-temperature phase, while the partition function computed on the dual 
lattice
$Z^{*}(K^{*})$ gives a good perturbative description of the low temperature
phase. 

From what we have seen up to now one may suspect that the equivalence
between totally different theories may only be a two-dimensional phenomenon.
This is actually not true because there is by now a overwhelming 
evidence~\cite{OSBO} that this also happens in the case of the 
four-dimensional supersymmetric Yang-Mills theory with gauge group $SU(2)$ and
with four supersymmetric conserved charges, the so-called $N=4$ supersymmetric
Yang-Mills theory. In general in this theory the original gauge symmetry is 
broken by the vacuum expectation value of some scalar fields in such a way 
that the gauge field corresponding to the unbroken $U(1)$ is left massless 
as the photon in the standard model, while all the other gauge fields, that 
we denote $W^{\pm}$, in analogy with the $W$-boson of the standard model, 
become 
massive. This theory has also soliton solutions that are monopoles and dyons 
with respect to the unbroken $U(1)$.  
It turns out that this theory with gauge coupling constant $g$ and 
with the particle content that we have
described above is exactly quantum equivalent to a supersymmetric $N=4$ 
Yang-Mills theory with gauge coupling constant $\hbar/ g$ where the 
fundamental fields are now the ones corresponding to the monopoles together 
with the dual, in the sense of the electromagnetic duality, of the massless 
$U(1)$ gauge fields realizing the beautiful Montonen-Olive duality 
idea~\cite{MONTOLI}. In this last theory the particles
corresponding to the massive fields $W^{\pm}$ appear instead as soliton 
solutions.

In conclusion the message that we get from field theory is that two theories
can appear to be very different from each other as far as their classical
Lagrangians or their respective weak coupling perturbation theories concern, 
but they can turn out to be completely quantum equivalent if we include also 
the non-perturbative effects. They are said to be dual to each other under 
weak-strong coupling duality that acts on the gauge coupling constant as 
$g \rightarrow  \hbar/g$.
Notice that the existence of the Planck constant is essential to match the
dimension of the two sides of the previous equation.
In particular we see that the fundamental particles that are those whose 
fields are present in the Lagrangian of a theory
appear as soliton solutions in the dual theory.

These ideas brought several people to investigate the soliton solutions in
various string theories or more precisely in the low energy limit of them
corresponding to the various supergravity theories. 
Through this investigation one has found classical "soliton" 
solutions corresponding to $p$-dimensional objects called $p$-branes, where 
$p=0$ for a point particle, $p=1$ for a string and so on, of various 
types~\footnote{For a review of the soliton solutions see Ref.~\cite{DUFF}.}. 
They are
\begin{enumerate}
\item{Smooth $p$-dimensional solitons with a mass per unit $p$-volume
proportional to the inverse of the square of the string coupling constant 
$g_{s}^{2}$. Those are the generalization to a
string theory of the solitons found in field theory}
\item{New types of $p$-dimensional objects, that are called $D$-branes,
having a mass per unit of $p$-volume proportional to the inverse of the
string coupling constant $g_s$. Those are new types of extended objects
that are not present in a field theory.}
\item{Extended $p$-dimensional objects corresponding to singular black holes.}
\end{enumerate}

Thus we find that, although we started from a string theory that by definition 
includes only one-dimensional objects, the theory contains also other
kinds of extended objects that, however, become infinitely heavy when $g_s 
\rightarrow 0$ disappearing from the spectrum. Our  string theory is 
therefore a pure string theory only in the perturbative regime 
$( g_s \rightarrow 0) $! 

We could ask ourselves if, in analogy with what we have learnt in several
examples in field theory, we could expect that our original string theory is
equivalent to another theory in which the $p$-branes appear as fundamental
states. Is there a duality between a string theory and another
string theory or more in general between a string theory and a $p$-brane 
theory?
 
In analogy with a particle or a string a $p$-brane is described by  its
space-time coordinate $x^{\mu} (\tau, \sigma_1, \sigma_2, \dots, \sigma_p )$
that is a function of the world-volume coordinates $\sigma_{i}$ of the 
$p$-brane. 
But, unlike particles and strings, we are not able to consistently quantize
$p$-branes with $p>1$. It is true that a $p$-brane as a string softens the
Minkowski space-time divergences, but its world-volume description has now 
so many degrees of freedom that we have not succeded in constructing a 
meaningful quantum theory of it.

But do we need to have a consistent quantum theory of an elementary arbitrary
$p$-brane? Or in other words does there exist a theory dual to one of the five 
inequivalent string theories where the fundamental elementary object is a
$p$-brane with $p>1$?

A detailed analysis~\cite{HULL} shows that there is no real need for it. 
In fact two of the five ten-dimensional consistent string 
theories, the type I and the heterotic one both with gauge group $SO(32)$, are
dual to each other in the sense of the weak-strong 
duality~\cite{WITTEN,OTHERS}, 
while a third string theory, the type IIB, is dual to itself, i.e. is 
self-dual~\cite{SCHWARZ}. Finally the other 
two theories, the type IIA and the heterotic one with gauge group $E_8 \otimes
E_8$, are dual to an $11$-dimensional particle theory~\cite{WITTEN,WITTEN2}. 
Actually their 
perturbative regime ($g_s \rightarrow 0$) can be obtained from an 
$11$-dimensional theory with the $11$-th dimension  compactified respectively
on a circle of radius $R$ or on a circle with opposite points identified 
(corresponding to the orbifold $S^{1}/Z_{2}$) in the limit in 
which the radius $R$, that in the ten-dimensional string metric is proportional
to a power $g_{s}^{2/3}$ of the string coupling constant, tends to zero.
This $11$-dimensional theory, called M-theory, has the
property of reducing at low energy to a unique $11$-dimensional supergravity
theory. It has also been shown that it  reduces to the other three string
theories in suitable limits~\cite{WITTEN2}.

In conclusion there is a strong evidence for the existence of a unique theory
that in certain specific limits reduces to one of the five consistent string 
theories in ten dimensions
and in these limits it can be given a very precise quantum meaning in 
perturbation theory in terms of the string coupling constant, but it is 
otherwise not a pure theory of strings at all.

It has been recently conjectured~\cite{BANKS} that the M-theory can be 
formulated as the $N$ going to infinity limit of the supersymmetric $N
\times N$
matrix quantum mechanics describing D0-branes. During the last half a year 
several
checks have confirmed this conjecture. The next few years will tell us if we
have finally constructed a useful formulation of a unique theory of all
forces.

Before concluding let us mention two very interesting applications of the
nonperturbative ideas developed above. 

The first concerns the unification
of the couplings. In sect. 2 we have discussed a problem concerning the
unifications of the couplings in the framework of the perturbative heterotic
string. However, from the value of the coupling constant at the unification
scale in eq. (\ref{gut}) one gets a value of $g_s \sim 1$ that is not 
consistent with the use of perturbation theory. But, if we go away from the
perturbative regime, as we have discussed above, one begins to see the $11$th
dimension and as a consequence the physics is approximately five-dimensional
already below the unification scale~\cite{WITTEN3}. In the $E_8 \otimes E_8$
heterotic string the gauge fields live each on one of the two ten-dimensional
walls of the $11$-dimensional world, while gravity lives in the entire 
$11$-dimensional world. This means that, below the unification scale where the
world is approximately five-dimensional, the dimensionless Planck constant is
$G_N E^3$ and not $G_N E^2$, as in four dimensions, while the gauge coupling 
constants still run according to a four-dimensional world. This makes the 
Newton's
constant run faster than before and consequently meet the gauge couplings at
a lower energy than before. It has been shown in Ref.~\cite{WITTEN3} that this
mechanism can cure the problem encountered in the perturbative heterotic 
string. 

The second concerns the calculation of the entropy of a black hole. It is 
known, since long time, that  black holes obey laws directly analogous
to the law of thermodynamics. In particular, they have a temperature and an
entropy given by $S = A/( 4 \hbar c G_{N})$, where $A$ is the area of the
horizon. This implies that in a statistical mechanics description of a
black hole one must identify the microstates that are responsible for the
entropy. It has been recently possible to identify them~\cite{VAFA} in the 
black holes
appearing in string theories and to show that their number correctly 
reproduces the entropy given by the Bekenstein-Hawking formula in terms of the
area of the horizon. 
   
To conclude, as few years before the fall of the Berlin wall it seemed 
impossible to believe that this event would have been possible in the next 
few years to come, so few years ago it seemed impossible that we could
understand in an analytical way so many aspects of non-perturbative field and
string theory and we could end up having a strong evidence for the
existence and even a conjectured formulation of a unique theory unifying all 
forces observed in nature.

On the other hand, as the fall of the Berlin wall did not solve all the
problems of the world, so we have a long way to go for understanding this
unique theory and especially to see if it will be able to predict what
is observed in the high energy experiments where smaller and smaller
structures are investigated.


\begin{thebibliography}{99}


\bibitem{LANDAU} L.D. Landau and E.M. Lifshitz, "The classical theory of 
           fields", Addison-Wesley Publishing Company, p. 97.

\bibitem{AMALDI} U. Amaldi, W. de Boer and H. F{\"{u}}rstenau, 
              {\it Phys. Lett.} {\bf 260B} (1991) 447.

\bibitem{SS} J. Scherk and J. Schwarz,  
            {\it Nucl. Phys.} {\bf B81} (1974) 118.

\bibitem{GSW} M. Green, J. Schwarz and E. Witten, "Superstring theory", 
              Cambridge University Press (1987).

\bibitem{GINSPARG} P. Ginsparg, {\it Phys. Lett.} {\bf 197B} (1987) 153.

\bibitem{KAPLU} V.S. Kaplunovsky, {\it Nucl. Phys.} {\bf B307} (1988) 145;
 Erratum: {\it ibid.} {\bf B382} (1992) 436.

\bibitem{DIENES} K.R. Dienes, {\it Nucl. Phys. Proc. Suppl.} {\bf 52A} (1997)
276.

\bibitem{COLE} A. Luther and I. Peschel, 
              {\it Phys. Rev.} {\bf B9} (1974) 2911;
              S. Coleman, {\it Phys. Rev.} {\bf D11} (1975) 2088;
              S. Mandelstam, {\it Phys. Rev.} {\bf D11} (1975) 3026.

\bibitem{OSBO} H. Osborn, {\it Phys. Lett.} {\bf 83B} (1979) 321;
               A. Sen, {\it Phys. Lett} {\bf 329B} (1994)  217;
               D.I. Olive, "Exact electromagnetic duality", hep/th 9508089.
\bibitem{MONTOLI} C. Montonen and D. Olive, 
                {\it Phys. Lett.} {\bf 72B} (1977) 117.

\bibitem{DUFF}  M.J. Duff, Ramzi R. Khuri and J.X. Lu, 
               {\it Phys. Rep.} {\bf 259} (1995) 326.

\bibitem{HULL} C.M. Hull, {\it Nucl. Phys.} {\bf B468} (1996) 113.

\bibitem{WITTEN} E. Witten, {\it Nucl. Phys. } {\bf B443} (1995) 85.

\bibitem{OTHERS} A. Dabholkar, {\it Phys. Lett} {\bf 357B} (1995) 307;
               C.M. Hull, {\it Phys. Lett} {\bf 357B} (1995) 545;
               J. Polchinsky and E. Witten, 
                 {\it Nucl. Phys.} {\bf B460} (1996) 525. 

\bibitem{SCHWARZ} C.M. Hull and P. Townsend, {\it Nucl. Phys.} {\bf B438}
                 (1995) 109.

\bibitem{WITTEN2} P. Horava and E. Witten, 
                   {\it Nucl. Phys.} {\bf B460} (1996) 506. 

\bibitem{BANKS} T. Banks, W. Fischler, S.H. Shenker and L. Susskind, {\it 
Phys. Rev.} {\bf D55} (1997) 5112. 

\bibitem{WITTEN3} E. Witten, {\it Nucl. Phys.} {\bf B471} (1996) 135. 

\bibitem{VAFA} C. Vafa and A. Strominger, {\it Phys. Lett.} {\bf 379B}
(1996) 99.

\end{thebibliography}
\end{document}